**Confidence Intervals for the $F_1$ Score:**
**A Comparison of Four Methods**

Kevin F. Y. Lam[1], Vikneswaran Gopal[2], Jiang Qian[1]

[1] Office of the President, National University of Singapore, Singapore
[2] Department of Statistics & Data Science, National University of Singapore, Singapore

**Author Note**

Lam Fu Yuan, Kevin  https://orcid.org/0000-0002-1535-7059
Jiang Qian  https://orcid.org/0009-0004-7029-6879

The authors have no relevant financial or non-financial interests to disclose.
Correspondence concerning this article should be addressed to
Lam Fu Yuan, Kevin, Office of the President, National University of Singapore.
Email: lamfy@nus.edu.sg

**Abstract**

In Natural Language Processing (NLP), binary classification algorithms are often evaluated using the $F_1$ score. Because the sample $F_1$ score is an estimate of the population $F_1$ score, it is not sufficient to report the sample $F_1$ score without an indication of how accurate it is. Confidence intervals are an indication of how accurate the sample $F_1$ score is. However, most studies either do not report them or report them using methods that demonstrate poor statistical properties. In the present study, we review current analytical methods (i.e., Clopper-Pearson method and Wald method) to construct confidence intervals for the population $F_1$ score, propose two new analytical methods (i.e., Wilson direct method and Wilson indirect method) to do so, and compare these methods based on their coverage probabilities and interval lengths, as well as whether these methods suffer from overshoot and degeneracy. Theoretical results demonstrate that both proposed methods do not suffer from overshoot and degeneracy. Experimental results suggest that both proposed methods perform better, as compared to current methods, in terms of coverage probabilities and interval lengths. we illustrate both current and proposed methods on a suggestion mining tasks. We discuss the practical implications of these results, and suggest areas for future research.

# 1        Introduction

## 1.1        Background

Natural Language Processing (NLP) is a subfield of computer science which uses computational techniques to learn, understand and produce human language content (Hirschberg & Manning, 2015). In NLP, the computational techniques are used to address problems which include, but are not limited to, supervised learning (Einstein, 2019).

In supervised learning, all of the data are labelled: the problem is a regression problem if the label is quantitative; and it is a classification problem if the label is qualitative (James et al., 2013). In a classification problem, the labels can include two (i.e., binary classification problem) or more (i.e., multi-class classification problem) categories (Jurafsky & Martin, 2023).

Regardless of whether the problem is a regression problem or a classification problem, the data are often split into a training set, a validation set and a test set: the training set is used to train the model, the validation set is used to tune the hyperparameters in the model, and the test set is used to evaluate the model (Jurafsky & Martin, 2023).

In a binary classification problem, some metrics used to evaluate the performance of the model are accuracy, precision, recall and the $F_1$ score (Jurafsky & Martin, 2023). However, although accuracy might seem to be a natural metric, it is seldom used in NLP because it does not perform well if the categories are imbalanced. Instead, precision, recall and the $F_1$ score, which is a single metric that incorporates both precision and recall, are preferred (Jurafsky & Martin, 2023; Takahashi et al., 2022; van Rijsbergen, 1975). In fact, the $F_1$ score has been used to evaluate the most recent developments in NLP, including Large Language Models (LLMs), from the Bidirectional Encoder Representations from Transformers (BERT) by Google (Devlin et al., 2019) to Large Language Model Meta AI-2 (LLaMA-2) by Meta (Touvron et al., 2023). Given both its prevalence in and its relevance to NLP, the $F_1$ score will be the focus of the present paper.

## 1.2        Problem

Regardless of the metric that is used to evaluate the performance of a model on the test set, it is assumed that both the observations in the test set and the observations in the future are drawn from the same distribution (Russel & Norvig, 2022). In other words, the test set is a sample from a population, and the metrics (e.g., accuracy, precision, recall, $F_1$ score) are sample estimates of the corresponding population parameters. If so, then it is not sufficient to report the sample estimate of the population parameter without an indication of how accurate it is (Lohr, 2022).

Confidence intervals are an indication of how accurate a sample estimate is (Lohr, 2022). However, most studies do not report confidence intervals for the population $F_1$ score, both in NLP (Wang et al., 2015) and elsewhere (Flores et al., 2022; Takahashi et al., 2022). Even among the studies that do, the confidence intervals either have coverage probabilities that are far from the nominal confidence level, have long interval lengths, suffer from either overshoot or degeneracy or are computationally intensive (Brown et al., 2001; Flores et al., 2022; Goutte & Gaussier, 2005; Newcombe, 1998; Park et al., 2023; Wang et al., 2015).

## 1.3    Contribution

Given the limitations of current methods to construct confidence intervals for the population $F_1$ score, we propose two analytical methods to do so. In the process, we answer the following three research questions:

**Research Question 1:** What are the current analytical methods to construct confidence intervals for the population $F_1$ score?

**Research Question 2:** What are proposed analytical methods (i.e., Wilson direct method and Wilson indirect method) to construct confidence intervals for the population $F_1$ score?

**Research Question 3:** How do the Wilson direct method and the Wilson indirect method perform, as compared to the current analytical methods, to construct confidence intervals for the population $F_1$ score?

## 1.4    Outline

The outline of the present paper is as follows: In Section II, we review the literature to define the $F_1$ score, to define the $F^*$ score and to describe the relationship between the $F_1$ score and the $F^*$ score. In Section III, we review the literature to define the sample $F_1$ score, to define the sample $F^*$ score, to demonstrate that the sample $F_1$ score is the maximum likelihood estimator of the population $F_1$ score, and to derive the asymptotic distribution of the sample $F_1$ score.

In Section IV, we review the literature to describe the current analytical methods to construct confidence intervals for the population $F_1$ score, describe the proposed analytical methods to construct confidence intervals for the population $F_1$ score, and prove that the proposed methods do not suffer from overshoot and degeneracy.

In Section V, we perform a simulation study to compare the confidence intervals constructed using the Clopper-Pearson method, the Wald method, the Wilson direct method and the Wilson indirect method, across the different simulation conditions, based on different evaluation criteria.

In Section VI, we illustrate the Clopper-Pearson method, the Wald method, the Wilson direct method and the Wilson indirect method, to construct confidence intervals for the population $F_1$ score, to evaluate the performance of Bidirectional Encoder Representations from Transformers (BERT) on two suggestion mining tasks using a public dataset.

In Section VII, we discuss the theoretical and practical implications of the study, and suggest directions for future research. In Section VIII, we conclude the present study.

## 2        $F_1$ Score

The $F_1$ score is the weighted harmonic mean of precision and recall in which both precision and recall are given equal weights (Jurafsky & Martin, 2023; van Rijsbergen, 1975):

$$F_1 = \left( \frac{\text{precision}^{-1} + \text{recall}^{-1}}{2} \right)^{-1} \tag{1}$$

In the literature, the $F_1$ score is also known as Sørensen-Dice coefficient (Dice, 1945; Sorensen, 1948). In this section, we review the literature to define the $F_1$ score, to define the $F^*$ score and to describe the relationship between the $F_1$ score and the $F^*$ score.

### 2.1        $F_1$ Score

Flores et al. (2022) demonstrated that the $F_1$ score can be stated in terms of either unconditional probabilities or conditional probabilities.

#### 2.1.1        Unconditional Probabilities

Let the unconditional probabilities $\boldsymbol{p} = (p_{11}, p_{10}, p_{01}, p_{00})$, where $p_{11}$ is the proportion of true positives, $p_{10}$ is the proportion of false positives, $p_{01}$ is the proportion of false negatives and $p_{00}$ is the proportion of true negatives among all documents. Table 1 summarises the unconditional probabilities $\boldsymbol{p}$ in a confusion matrix.

[Insert Table 1 Here.]

The unconditional probabilities $\boldsymbol{p}$ can be used to obtain both precision and recall.

Precision, also known as the positive predictive value (Rosner, n.d.), is the proportion of true positives among all documents that are predicted positives (Jurafsky & Martin, 2023):

$$\text{precision} = \frac{p_{11}}{p_{11} + p_{10}} \tag{2}$$

Recall, also known as sensitivity (Rosner, n.d.), is the proportion of true positives among all documents that are actual positives (Jurafsky & Martin, 2023).

$$\text{recall} = \frac{p_{11}}{p_{11} + p_{01}} \tag{3}$$

Then the $F_1$ score can be stated in terms of the unconditional probabilities $\boldsymbol{p}$:

$$F_1 = \left(\frac{\text{precision}^{-1} + \text{recall}^{-1}}{2}\right)^{-1} \tag{4}$$

$$= \left(\frac{1}{2} \cdot \frac{p_{11} + p_{10}}{p_{11}} + \frac{1}{2} \cdot \frac{p_{11} + p_{01}}{p_{11}}\right)^{-1} \tag{5}$$

$$= \frac{2p_{11}}{2p_{11} + p_{10} + p_{01}} \tag{6}$$

### 2.1.2 Conditional Probabilities

Let the conditional probabilities $\boldsymbol{\pi} = (\pi_{11}, \pi_{10}, \pi_{10})$, where $\pi_{11} = p_{11}/(p_{11} + p_{10} + p_{01})$ is the proportion of true positives, $\pi_{10} = p_{10}/(p_{11} + p_{10} + p_{01})$ is the proportion of false positives, $\pi_{01} = p_{01}/(p_{11} + p_{10} + p_{01})$ is the proportion of false negatives, among all relevant documents (i.e., all documents that are either actual positives or predicted positives). Table 2 summarises the conditional probabilities $\boldsymbol{\pi}$ in a confusion matrix.

[Insert Table 2 Here.]

Then the $F_1$ score can also be stated in terms of the conditional probabilities $\boldsymbol{\pi}$:

$$F_1 = \frac{2p_{11}}{2p_{11} + p_{10} + p_{01}} \tag{7}$$

$$= \frac{2\pi_{11}(p_{11} + p_{10} + p_{01})}{(2\pi_{11} + \pi_{10} + \pi_{01})(p_{11} + p_{10} + p_{01})} \tag{8}$$

$$= \frac{2\pi_{11}}{1 + \pi_{11}} \tag{9}$$

## 2.2 $F^*$ Score

Hand et al. (2021) termed $\pi_{11}$ as the $F^*$ score, also known as the Jaccard coefficient (Jaccard, 1912). In other words, the $F_1$ score can be stated in terms of the $F^*$ score:

$$F_1 = \frac{2F^*}{1 + F^*} \tag{10}$$

## 2.3 $F_1$ Score and $F^*$ Score

Hand *et al.* (2021) demonstrated that the $F_1$ score is a monotonic function of the $F^*$ score on the interval [0,1].

**Proposition 1.** *The $F_1$ score is a monotonic function of the $F^*$ score on the interval* [0,1].

*Proof.* From (10), the first derivative of $F_1$ with respect to $F^*$ is as follows:

$$\frac{\partial F_1}{\partial F^*} = \frac{2}{1 + F^*} - \frac{2F^*}{(1 + F^*)^2} \tag{11}$$

$$= \frac{2}{(1 + F^*)^2} \tag{12}$$

By the First Derivative Test (e.g., Hass et al., 2024, p. 250), because $\frac{\partial F_1}{\partial F^*} > 0$ at each point $F^* \in (0,1)$, $F_1$ is increasing on $[0,1]$. Therefore, the $F_1$ score is a monotonic function of the $F^*$ score on the interval $[0,1]$. Figure 1 illustrates the relationship between the $F_1$ score and the $F^*$ score on the interval $[0,1]$.                   □

[Insert Figure 1 Here.]

## 3      Sample $F_1$ Score

In this section, we review the literature to define the sample $F_1$ score, to define the sample $F^*$ score, to demonstrate that the sample $F_1$ score is the maximum likelihood estimator of the population $F_1$ score, and to derive the asymptotic distribution of the sample $F_1$ score.

### 3.1      Sample $F_1$ Score

Flores et al. (Flores et al., 2022) demonstrated that the sample $F_1$ score can be stated in terms of either unconditional probabilities or conditional probabilities.

#### 3.1.1      Unconditional Probabilities

Let $\boldsymbol{n} = (n_{11}, n_{10}, n_{01}, n_{00})$, where $n_{11}$ is the number of true positives, $n_{10}$ is the number of false positives, $n_{01}$ is the number of false negatives, and $n_{00}$ is the number of true negatives, among all $n$ documents in a sample.

Then $\boldsymbol{n}$ can be assumed to follow a multinomial distribution with parameters $n$ and $\boldsymbol{p}$ (Flores et al., 2022; Goutte & Gaussier, 2005; Shang et al., 2023; Takahashi et al., 2022):

$$\boldsymbol{n} \sim Multinomial(n; \boldsymbol{p})$$

Both $n$ and $\boldsymbol{n}$ can be used to obtain the unconditional probabilities $\hat{\boldsymbol{p}} = (\hat{p}_{11}, \hat{p}_{10}, \hat{p}_{01}, \hat{p}_{00})$, where $\hat{p}_{11} = n_{11}/n$ is the proportion of true positives, $\hat{p}_{10} = n_{10}/n$ is the proportion of false positives, $\hat{p}_{01} = n_{01}/n$ is the proportion of false negatives and $\hat{p}_{00} = n_{00}/n$ is the proportion of true negatives, among all $n$ documents in the sample.

Then the sample $F_1$ score can be stated in terms of the unconditional probabilities $\hat{\boldsymbol{p}}$:

$$\hat{F}_1 = \frac{2\hat{p}_{11}}{2\hat{p}_{11} + \hat{p}_{10} + \hat{p}_{01}} \tag{14}$$

### 3.1.2 Conditional Probabilities

Let $\boldsymbol{v} = (n_{11}, n_{10}, n_{01})$, where $n_{11}$ is the number of true positives, $n_{10}$ is the number of false positives, and $n_{01}$ is the number of false negatives, among all $n = v + n_{00}$ documents in a sample.

Then $v$ can be assumed to follow a binomial distribution with parameters $n$ and $p_{11} + p_{10} + p_{01}$ (Flores et al., 2022):

$$v \sim Binomial(n, p_{11} + p_{10} + p_{01}) \tag{15}$$

And $n_{11}$ can be assumed to follow a binomial distribution with parameters $v$ and $F^*$, conditioned on the observed value of $v$ (Flores et al., 2022):

$$n_{11} \sim Binomial(v, F^*) \tag{16}$$

Both $v$ and $\boldsymbol{v}$ can be used to obtain the conditional probabilities $\hat{\boldsymbol{\pi}} = (\hat{\pi}_{11}, \hat{\pi}_{10}, \hat{\pi}_{01})$, where $\hat{\pi}_{11} = n_{11}/v$ is the proportion of true positives, $\hat{\pi}_{10} = n_{10}/v$ is the proportion of false positives, and $\hat{\pi}_{01} = n_{01}/v$ is the proportion of false negatives, among all $v$ relevant documents in the sample.

Then the sample $F_1$ score can also be stated in terms of the conditional probabilities $\hat{\boldsymbol{\pi}}$:

$$\hat{F}_1 = \frac{2\hat{\pi}_{11}}{1 + \hat{\pi}_{11}} \tag{17}$$

### 3.2 Sample $F^*$ Score

If $\hat{\pi}_{11}$ is termed the sample $F^*$ score, then the sample $F_1$ score can also be stated in terms of the sample $F^*$ score:

$$\hat{F}_1 = \frac{2\hat{F}^*}{1 + \hat{F}^*} \tag{18}$$

### 3.3 Maximum Likelihood Estimation of the Population $F_1$ Score

Regardless of whether the sample $F_1$ score is stated in terms of the unconditional probabilities $\hat{\boldsymbol{p}}$ or the conditional probabilities $\hat{\boldsymbol{\pi}}$ (i.e., the sample $F^*$ score), the sample $F_1$ score is the maximum likelihood estimator of the population $F_1$ score.

**Proposition 2.** *The sample $F_1$ score is the maximum likelihood estimator of the population $F_1$ score.*

*Proof.* By the invariance property of maximum likelihood estimators (e.g., Casella & Berger, 2002, p. 320), because the sample proportion (i.e., $\hat{F}^*$) is the maximum likelihood estimator of the population proportion (i.e., $F^*$) (e.g., Casella & Berger, 2002, p. 318), and $F_1$ is a one-to-one function of $F^*$ on the interval $[0,1]$ (Proposition 1), sample $F_1$ score is also the maximum likelihood estimator of the population $F_1$ score. □

### 3.4 Asymptotic Distribution of the Sample $F_1$ Score

Flores et al. (2022) used the Delta Method to derive the asymptotic distribution of the $F_1$ score.

**Proposition 3.** *If $\nu$ is sufficiently large, then $(\hat{F}_1 - F_1)/\sigma_{\hat{F}_1}$ has approximately the standard normal distribution.*

*Proof.* By the Central Limit Theorem (e.g., Sen & Singer, 2018, p. 107), or the De Moivre-Laplace Theorem (e.g., Sen & Singer, 2018, p. 108), $(\hat{F}^* - F^*)/\sigma_{\hat{F}^*}$ converges in distribution to the standard normal distribution:

$$\frac{(\hat{F}^* - F^*)}{\sigma_{\hat{F}^*}} \xrightarrow{D} N(0,1) \tag{19}$$

where $\sigma_{\hat{F}^*}^2$ is the variance of $\hat{F}^*$:

$$\sigma_{\hat{F}^*}^2 = \frac{F^*(1 - F^*)}{\nu} \tag{20}$$

By the Delta Method (e.g., Sen & Singer, 2018, p. 131; Stone, 1995, p. 637), $(\hat{F}_1 - F_1)/\sigma_{\hat{F}_1}$ also converges in distribution to the standard normal distribution:

$$\frac{(\hat{F}_1 - F_1)}{\sigma_{\hat{F}_1}} \xrightarrow{D} N(0,1) \tag{21}$$

where $\sigma_{\hat{F}_1}^2$ is the variance of $\hat{F}_1$:

$$\sigma_{\hat{F}_1}^2 = \left[\frac{\partial F_1}{\partial F^*}\right]^2 \frac{\sigma_{F^*}^2}{\nu} \tag{22}$$

$$= \left[\frac{2}{(1 + F^*)^2}\right]^2 \frac{F^*(1 - F^*)}{\nu} \tag{23}$$

$$= \frac{4F^*(1 - F^*)}{\nu(1 + F^*)^4} \tag{24}$$

$$= \frac{F_1(1 - F_1)(2 - F_1)^2}{2\nu} \tag{25}$$

Therefore, if $\nu$ is sufficiently large, then $(\hat{F}_1 - F_1)/\sigma_{\hat{F}_1}$ has approximately the standard normal distribution. □

## 4        Confidence Intervals for the $F_1$ Score

A $(1 - \alpha)100\%$ confidence interval for a population parameter, where $\alpha$ is the level of statistical significance (e.g., 0.05), is the expected percentage of intervals that include the true population parameter if repeated samples from the population are taken, and a confidence interval is constructed using the same method for each possible sample (e.g., Lohr, 2022).

In this section, we review the literature to describe the current analytical methods to construct confidence intervals for the population $F_1$ score (i.e., Clopper-Pearson method and Wald method), describe the proposed analytical methods to construct confidence intervals for the population $F_1$ score (i.e., Wilson direct method and Wilson indirect method), and prove that the proposed methods do not suffer from both overshoot and degeneracy.

### 4.1        Clopper-Pearson Method

The Clopper-Pearson method assumes that $n_{11}$ has a binomial distribution with parameters $\nu$ and $F^*$ (16) and inverts the binomial test for the sample $F^*$ score (Brown et al., 2001).

The endpoints of the confidence interval for the population $F^*$ score are the solutions in $F^*$ to the following equations:

$$P_{F^*}(n_{11} < \tilde{n}_{11}) = 1 - \alpha/2 \tag{26}$$

and

$$P_{F^*}(n_{11} > \tilde{n}_{11}) = 1 - \alpha/2 \tag{27}$$

where $\tilde{n}_{11}$ is the realisation of $n_{11}$.

In particular, the lower endpoint is the $\alpha/2$ quantile of a beta distribution with parameters $\tilde{n}_{11}$ and $\nu - \tilde{n}_{11} + 1$, and the upper endpoint is the $1 - \alpha/2$ quantile of a beta distribution with parameters $\tilde{n}_{11} + 1$ and $\nu - \tilde{n}_{11}$ (e.g., Brown et al., 2001).

The endpoints of the confidence interval for the population $F_1$ score are the abovementioned solutions' transformation via (10).

### 4.2        Wald Method

The Wald method assumes that $(\hat{F}_1 - F_1)/\sigma_{\hat{F}_1}$ has approximately a standard normal distribution (Proposition 3) and  inverts the Wald test for the sample $F_1$ score (Brown et al., 2001):

$$\left|\frac{\hat{F}_1 - F_1}{\hat{\sigma}_{\hat{F}_1}}\right| < z_{\alpha/2} \tag{28}$$

Where $z_{\alpha/2} = \Phi^{-1}(1 - \alpha/2)$, $\Phi(z)$ is the standard normal distribution function, and $\hat{\sigma}_{\hat{F}_1}^2$ is the estimated variance of the sample $F_1$ score using the maximum likelihood estimator of the population $F_1$ score (Proposition 2):

$$\hat{\sigma}_{\hat{F}_1}^2 = \frac{\hat{F}_1(1 - \hat{F}_1)(2 - \hat{F}_1)^2}{2\nu} \tag{29}$$

The endpoints of the confidence interval for the population $F_1$ score using the Wald method are the solutions in $F_1$ to the following equations:

$$F_1 = \hat{F}_1 - z_{\alpha/2} \times \hat{\sigma}_{\hat{F}_1} \tag{30}$$

and

$$F_1 = \hat{F}_1 + z_{\alpha/2} \times \hat{\sigma}_{\hat{F}_1} \tag{31}$$

### 4.3    Wilson Direct Method

The Wilson direct method also assumes that $(\hat{F}_1 - F_1)/\sigma_{\hat{F}_1}$ has approximately a standard normal distribution (Proposition 3). However, it uses the null variance, as provided in (25), instead of the estimated variance, as provided in (29), when inverting the score test for the sample $F_1$ score (Brown et al., 2001):

$$\left|\frac{\hat{F}_1 - F_1}{\sigma_{\hat{F}_1}}\right| < z_{\alpha/2} \tag{32}$$

The endpoints of the confidence interval for the population $F_1$ score are the real (i.e., not imaginary) solutions in $F_1$ to the following quartic equation, most conveniently solved by iteration (e.g., Jenkins & Traub, 1970):

$$\left(\frac{\hat{F}_1 - F_1}{\sigma_{\hat{F}_1}}\right)^2 = z_{\alpha/2}^2 \tag{33}$$

$$(\hat{F}_1 - F_1)^2 = z_{\alpha/2}^2 \frac{F_1(1 - F_1)(2 - F_1)^2}{2\nu} \tag{34}$$

If $k = z_{\alpha/2}^2/\nu$, then

$$2F_1^2 - 4\hat{F}_1 F_1 + 2\hat{F}_1^2 = k(F_1 - F_1^2)(4 - 4F_1 + F_1^2) \tag{35}$$

$$2F_1^2 - 4\hat{F}_1 F_1 + 2\hat{F}_1^2 = k(4F_1 - 8F_1^2 + 5F_1^3 - F_1^4) \tag{36}$$

$$kF_1^4 - 5kF_1^3 + 2(4k+1)F_1^2 - 4(k + \hat{F}_1)F_1 + 2\hat{F}_1^2 = 0 \tag{37}$$

### 4.4    Wilson Indirect Method

The Wilson indirect method assumes that $(\hat{F}^* - F^*)/\sigma_{\hat{F}^*}$ has approximately the standard normal distribution (19) and inverts the score test for the sample $F^*$ score:

$$\left| \frac{\hat{F}^* - F^*}{\sigma_{\hat{F}^*}} \right| < z_{\alpha/2} \tag{38}$$

where $\sigma_{\hat{F}^*}^2$ is the null variance of $\hat{F}^*$ as provided in (20).

The endpoints of the confidence interval for the population $F^*$ score using the Wilson method are the solutions in $F^*$ to the following quadratic equation:

$$\left( \frac{\hat{F}^* - F^*}{\sigma_{\hat{F}^*}} \right)^2 = z_{\alpha/2}^2 \tag{39}$$

$$\left( \hat{F}^* - F^* \right)^2 = z_{\alpha/2}^2 \frac{F^*(1 - F^*)}{\nu} \tag{40}$$

If $k = z_{\alpha/2}^2 / \nu$, then

$$\hat{F}^{*2} - 2\hat{F}^* F^* + F^{*2} = kF^* - kF^{*2} \tag{41}$$

$$(1 + k)F^{*2} - (2\hat{F}^* + k)F^* + \hat{F}^{*2} = 0 \tag{42}$$

The endpoints of the confidence interval for the population $F_1$ score using the Wilson indirect method are the abovementioned solutions' transformation via (10).

### 4.5    Overshoot and Degeneracy

Because confidence intervals for the population $F_1$ score constructed using the Wald method produce intervals centred on the point estimate, these intervals suffer from both overshoot, in which either the upper limit is greater than 1 or the lower limit is less than 0, and degeneracy, in which the confidence interval has zero width (Flores et al., 2022; Newcombe, 1998; Takahashi et al., 2022). And because confidence intervals for the population $F^*$ score, constructed using either the Clopper-Pearson method or the Wilson score method, do not suffer from overshoot and degeneracy (Newcombe, 1998), those for the population $F_1$ score, constructed using either the Clopper-Pearson method or the Wilson indirect method, obtained from a strictly monotonic transformation via (10), also do not suffer from overshoot and degeneracy.

However, the properties of confidence intervals constructed using the Wilson direct method are less apparent and have not been studied in the literature, to the best of the authors' knowledge. In the remainder of the section, we prove that confidence intervals for the population $F_1$ score constructed using the Wilson direct method also do not suffer from overshoot and degeneracy because they have exactly 2 distinct real roots under almost all conditions in practice (e.g., to have at least 3 relevant observations in the test set if the level of statistical significance is set at 0.05).

In the proof, let $f(F_1)$ be the quartic polynomial in the left-hand-side of (37). $f(F_1)$ can be further expressed as follows:

$$f(F_1) = kF_1^4 - 5kF_1^3 + 2(4k+1)F_1^2 - 4(k+\hat{F}_1)F_1 + 2\hat{F}_1^2 \qquad (43)$$

$$= kF_1(F_1-1)(F_1-2)^2 + 2(F_1-\hat{F}_1)^2 \qquad (44)$$

If so, then the first derivative is as follows:

$$f'(F_1) = \frac{\partial}{\partial F_1} f(F_1) \qquad (45)$$

$$= 4kF_1^3 - 15kF_1^2 + 16kF_1 + 4F_1 - 4(k+\hat{F}_1) \qquad (46)$$

$$= 4kF_1^3 - 15kF_1^2 + 16kF_1 + 4F_1 - 4k - 4\hat{F}_1 \qquad (47)$$

$$= 4kF_1^3 - 15kF_1^2 + 16kF_1 - 4k + 4(F_1-\hat{F}_1) \qquad (48)$$

$$= k(4F_1^3 - 15F_1^2 + 16F_1 - 4) + 4(F_1-\hat{F}_1) \qquad (49)$$

$$= k(F_1-2)(4F_1^2 - 7F_1 + 2) + 4(F_1-\hat{F}_1) \qquad (50)$$

And the second derivative is as follows:

$$f''(F_1) = \frac{\partial^2}{\partial F_1^2} f(F_1) \qquad (51)$$

$$= 12kF_1^2 - 30kF_1 + 16k + 4 \qquad (52)$$

**Lemma 1.** $f(F_1)$ *has at least 2 distinct real roots in the interval* $[0,1]$ *if* $k > 0$.

*Proof.* First, suppose that $\hat{F}_1 = 0$. Then $F_1 = 0$ is a root. By the Intermediate Value Theorem (e.g., Hass et al., 2024, p. 127), because $f'(0) < 0$ (i.e., $f(0+h) < 0$ for some small increment $h$) if $k > 0$, and $f(1) > 0$, $f(F_1)$ has a real root in the interval $(0,1)$ if $k > 0$.

Second, suppose that $0 < \hat{F}_1 < 1$. By the Intermediate Value Theorem, because $f(\hat{F}_1) < 0$ if $k > 0$, and $f(0) > 0$, $f(F_1)$ has a real root in the interval $(0, \hat{F}_1)$ if $k > 0$. And because $f(1) > 0$, $f(F_1)$ also has a real root in the interval $(\hat{F}_1, 1)$ if $k > 0$.

Last, suppose that $\hat{F}_1 = 1$. Then $F_1 = 1$ is a root. By the Intermediate Value Theorem, because $f'(1) > 0$ (i.e., $f(1 - h) < 0$ for some small decrement $h$) if $k > 0$, and $f(0) > 0$, $f(F_1)$ has a real root in the interval $(0,1)$ if $k > 0$.

Therefore, $f(F_1)$ has at least 2 distinct real roots in the interval [0,1] if $k > 0$.                            □

**Lemma 2.** $f(F_1)$ *has at most* 2 *distinct real roots if* $0 < k < 16/11$.

*Proof.* If $f(F_1)$ is concave up, then it has at most 2 distinct real roots. By the Second Derivative Test for Concavity (e.g., Hass et al., 2024, p. 256), $f(F_1)$ is concave up if $f''(F_1) > 0$. And $f''(F_1) > 0$ if it is also concave up (i.e., $12k > 0$) and its discriminant is strictly negative:

$$(-30k)^2 - 4(12k)(16k + 4) < 0 \tag{53}$$
$$132k^2 - 192k < 0 \tag{54}$$

Which implies that

$$0 < k < 16/11 \tag{55}$$

Therefore, $f(F_1)$ has at most 2 distinct real roots if $0 < k < 16/11$.                            □

**Theorem 1.** $f(F_1)$ *has exactly* 2 *distinct real roots in the interval [0,1] if* $0 < k < 16/11$.

*Proof.* The theorem follows from Lemma 1 and Lemma 2.                            □

Therefore, the confidence intervals for the population $F_1$ score constructed using the Wilson direct method do not suffer from overshoot and degeneracy if $v > (11/16) \times z_{a/2}^2$ and $z_{a/2}^2 > 0$. For example, if the level of statistical significance is 0.05, then the number of relevant observations in the test set should be at least 3.

## 5        Simulation Study

In this section, we perform a simulation study to compare the 95% confidence intervals for the population $F_1$ score constructed using the Clopper-Pearson method, the Wald method, the Wilson direct method and the Wilson indirect method, across the different simulation conditions, based on different evaluation criteria. In particular, we describe the simulation conditions, the evaluation criteria and the study results.

## 5.1 Simulation Conditions

The simulation conditions were adapted from Takahashi et al. (2022). In particular, 18 simulation conditions were obtained from crossing six $n$ (i.e., $25, 50, 100, 500, 1000, 5000$) against three scenarios:

**Scenario 1**: The positive class has moderate prevalence (50%), high precision (80%) and high recall (80%). Therefore, $p_{11} = 0.4$, $p_{10} = 0.1$, $p_{01} = 0.1$, $p_{00} = 0.4$ and $F_1 = 0.8$.

**Scenario 2**: The positive class has high prevalence (80%), high precision (80%) and high recall (80%). Therefore, $p_{11} = 0.64$, $p_{10} = 0.16$, $p_{01} = 0.16$, $p_{00} = 0.04$ and $F_1 = 0.8$.

**Scenario 3**: The positive class has high prevalence (80%), high precision (80%) and low recall (20%). Therefore, $p_{11} = 0.16$, $p_{10} = 0.04$, $p_{01} = 0.64$, $p_{00} = 0.16$ and $F_1 = 0.32$.

For each simulation condition, we generated $10^6$ replicates. Each replicate was generated from a multinomial distribution with the corresponding $n$, $p_{11}$, $p_{10}$, $p_{01}$ and $p_{00}$ for that condition. The generation of replicates from multinomial distributions have been applied in previous studies which examined the properties of confidence intervals for the $F_1$ score (Flores et al., 2022; Takahashi et al., 2022).

## 5.2 Evaluation Criteria

In each simulation condition, we evaluated each method based on their coverage probabilities, expected lengths, overshoot probabilities and degeneracy probabilities. The coverage probability is the proportion of times the confidence interval includes the true population parameter (Newcombe, 1998). The expected length is the expected difference between the upper and lower limits of the confidence interval (Brown et al., 2001). The overshoot probability is the proportion of times the confidence interval has either an upper limit that is greater than 1 or a lower limit that is less than 0 (Newcombe, 1998). The degeneracy probability is the proportion of times the confidence interval has zero width (i.e., the difference between the upper and lower limits of the confidence interval is zero; Newcombe, 1998). It is desirable to have confidence intervals with coverage probabilities near the nominal confidence level, with small average lengths, and that do not suffer from overshoot and degeneracy.

## 5.3 Study Results

### 5.3.1 Coverage Probabilities

For small $n$, across all scenarios, the Clopper-Pearson method demonstrated coverage probabilities that were far greater than the nominal confidence level, the Wald method demonstrated coverage probabilities that were far less than the nominal confidence level, and both the Ward direct method and the Ward indirect method demonstrated coverage probabilities that were near the nominal confidence level. For large $n$, across all scenarios, all four methods demonstrated coverage probabilities that were near the nominal confidence level. Table 3 and Figure 2A summarise the coverage probabilities for each method across all simulation conditions.

[Insert Table 3 Here]

[Insert Figure 2 Here]

### 5.3.2 Expected Lengths

For small $n$, in Scenarios 1 and 2, the Wilson indirect method demonstrated the shortest expected lengths as compared to the Clopper-Pearson method, the Wald method and the Wilson direct method; but in Scenario 3, the Wilson direct method demonstrated the shortest expected lengths as compared to the Clopper-Pearson method, the Wald method and the Wilson indirect method. For large $n$, across all scenarios, all four methods demonstrated comparable expected lengths. For each scenario, all methods' expected lengths decreased as $n$ increased, as expected. Table 4 and Figure 2B summarises the expected lengths for each method across all simulation conditions.

[Insert Table 4 Here.]

The difference in expected lengths between the confidence intervals constructed using the Wilson direct method and the Wilson indirect method occurs because both methods' interval lengths depend on both $v$ and $\hat{F}_1$. Figure 3 illustrates a comparison of the interval lengths between the Wilson direct method, obtained from the absolute difference between the roots to (35), and the Wilson indirect method, obtained from the absolute difference between the transformed roots to (42) (i.e., via (10)), across different $v$ and $\hat{F}_1$.

[Insert Figure 3 Here.]

If $v$ is small and $\hat{F}_1$ is small, then the Wilson direct method produces confidence intervals that are shorter as compared to the Wilson indirect method. However, if $v$ is small and $\hat{F}_1$ is large, then the Wilson indirect method produces confidence intervals that are shorter as compared to the Wilson direct method. And if $v$ is large, then both methods produce confidence intervals that are of comparable lengths.

For example, in Scenario 3 where $n = 25$ and $F_1 = 0.32$, the Wilson direct method demonstrated a shorter expected length (0.395) as compared to the Wilson indirect method (0.414) (Table 4). This occurred because $n$ was small, and therefore $v \leq n$ was small, and because the population $F_1$ score was small, and therefore $\hat{F}_1$ tended to be small.

### 5.3.3 Overshoot Probabilities

For small $n$, across all scenarios, the Wald method demonstrated overshoot. However, for large $n$, across all scenarios, the Wald method did not demonstrate overshoot. Regardless of $n$, across all scenarios, neither the Clopper-Pearson method, the Wilson direct method nor the Wilson indirect method demonstrated overshoot. Table 5 summarises the overshoot probabilities for each method across all simulation conditions.

[Insert Table 5 Here.]

### 5.3.4    Degeneracy Probabilities

For small $n$, especially in Scenarios 1 and 3, the Wald method demonstrated degeneracy. However, for large $n$, across all scenarios, the Wald method did not demonstrate degeneracy. Regardless of $n$, across all scenarios, neither the Clopper-Pearson method, the Wilson direct method nor the Wilson indirect method demonstrated degeneracy. Table 6 summarises the degeneracy probabilities for each method across all simulations conditions.

[Insert Table 6 Here.]

## 6       Example

In this section, we illustrate the Clopper-Pearson method, the Wald method, the Wilson direct method and the Wilson indirect method to construct confidence intervals for the $F_1$ score to evaluate the performance of Bidirectional Encoder Representations from Transformers (BERT) on two suggestion mining tasks using a public dataset.

On the one hand, suggestion mining is the automatic extraction of suggestions using techniques in NLP (Brun & Hagège, 2013). In suggestion mining, an explicit suggestion directly suggests or recommends an action or entity, and an implicit suggestion indirectly suggests or recommends an action or entity (Negi et al., 2018). In the literature, suggestion mining often focuses on explicit suggestions, and has been applied to extract suggestions across a number of industries including, but not limited to, education, software and tourism (e.g., Gottipati et al., 2018; Negi et al., 2019).

On the other hand, BERT is a language model built on a multi-layer bidirectional transformer encoder architecture based on the original implementation described in Vaswani et al. (2017). BERT was pre-trained on a masked language modelling task and a next sentence prediction task using data from BooksCorpus and English Wikipedia (Devlin et al., 2019). BERT_BASE comprises 12 layers (i.e., transformer blocks), 768 hidden units and 12 self-attention heads (i.e., 110 million parameters); and BERT_LARGE comprises 24 layers, 1024 hidden units and 16 attention heads (i.e., 340 million parameters).

In the present example, we used the fine-tuning approach on BERT_BASE in which a simple classification layer was added to the pre-trained model and all parameters were jointly fine-tuned on the corresponding downstream task (Devlin et al., 2019). As suggested by Devlin et al. (2019), we ran an exhaustive search over the following hyperparameters and chose the model that performed best on the validation set:

- **Batch size:** 16, 32

- **Learning rate (Adam):** 5e-5, 3e-5, 2e-5
- **Number of epochs:** 2, 3, 4

Yamamoto and Sekiya (2019) used the fine-tuning approach on BERT$_{\text{BASE}}$ and applied it to the Semantic Evaluation 2019 (SemEval2019) Task 9 Subtask A dataset (Negi et al., 2019). The dataset was scraped from a software forum. It was annotated in two phases. In Phase 1, crowdsourced annotators labelled all observations as either a suggestion or a non-suggestion. And in Phase 2, expert annotators labelled some samples, in particular only those labelled as a suggestion in Phase 1, as either a suggestion or a non-suggestion. In the final dataset, an observation was labelled as a suggestion if it was labelled as such in both Phase 1 and Phase 2. Otherwise, the observation was labelled as a non-suggestion.

The final dataset comprised 9,925 observations among which 2,468 (25%) were suggestions. The dataset was split into a training set, a validation set and a test set. The training set comprised 8,500 observations among which 2,085 (25%) were suggestions. The validation set comprised 592 (6%) observations among which 296 were suggestions (50%). And the test set comprised 833 (8%) observations among which 87 (10%) were suggestions.

In Yamamoto and Sekiya (2019), BERT$_{\text{BASE}}$ achieved an $F_1$ score of 0.731 on the test set. In the present example, BERT$_{\text{BASE}}$ achieved 77 true positives, 44 false positives, 10 false negatives and 702 true negatives on the test set. This corresponds to an $F_1$ score of 0.740.

Using the Clopper-Pearson method, the 95% confidence interval is [0.665,0.805] and has an interval length of 0.139. Using the Wald method, the 95% confidence interval is [0.674,0.807] and has an interval length of 0.134. Using the Wilson direct method, the 95% confidence interval is [0.664,0.799] and has an interval length of 0.135. And using the Wilson indirect method, the 95% confidence interval is [0.669,0.801] and has an interval length of 0.133.

As expected, the confidence interval constructed using the Clopper-Pearson method was the longest. The confidence interval constructed using the Wilson indirect method was shorter as compared to that constructed using the Wilson direct method because the sample $F_1$ score was relatively large.

## 7        Discussion

In the present paper, we reviewed the current analytical methods to construct confidence intervals for the population $F_1$ score, proposed two analytical methods to do so, and compared their performances. The answers to the three research questions formulated in Section 1 are as follows:

- First, the current analytical methods to construct the confidence interval for the population $F_1$ score include the Clopper-Pearson method, which inverts the Binomial test for the sample count, and the Wald method, which inverts the Wald test for the sample $F_1$ score.

- Second, the proposed analytical methods to construct confidence intervals for the population $F_1$ score include the Wilson direct method, which inverts the score test for the sample $F_1$ score, and the Wilson indirect method, which inverts the score test for the sample $F^*$ score.

- Last, both the Wilson direct method and the Wilson indirect method perform better, in terms of coverage probabilities and average lengths, as compared to the Clopper-Pearson method and the Wald method to construct confidence intervals for the population $F_1$ score. In addition, unlike the Wald method, neither the Wilson direct method nor the Wilson indirect method suffer from overshoot and degeneracy.

In accordance with these findings, Takahashi et al. (2022) also reported that the coverage probabilities for the Wald method to construct confidence intervals for the population $F_1$ score tended to be far smaller as compared to the nominal confidence level when $n < 100$ (p. 10).

Because the sample $F_1$ score is an estimate of the population $F_1$ score (Lohr, 2022), confidence intervals should be used to indicate how accurate the estimate is. In the construction of confidence intervals for the population $F_1$ score, if the test set is small (i.e., $n \leq 100$), then both the Wilson direct method and the Wilson indirect method are recommended, especially since both methods demonstrate better coverage probabilities and interval lengths as compared to the Clopper-Pearson method and the Wald method. Because both methods demonstrate comparable coverage probabilities for small $n$, the choice between them depends on the interval length. If the test set is large (i.e., $n > 100$), then both the Wilson direct method and the Wilson indirect method are also recommended, especially since both methods do not suffer from overshoot and degeneracy. Because both methods demonstrate comparable coverage probabilities and interval lengths for large $n$, the choice between them depends on individual preference.

The recommendation of methods which construct confidence intervals using the score test (i.e., Wilson direct method and Wilson indirect method), as compared to using either the binomial test (i.e., Clopper-Pearson method) or the Wald test (i.e., Wald method), is also consistent with the literature. In the construction of confidence intervals for the population proportion, Brown et al. (2001) recommended the Wilson method regardless of sample size. The authors also argued that the Clopper-Pearson method is "wastefully conservative and is not a good choice for practical use" (p. 113), and that the Wald method is "persistently chaotic and unacceptably poor" (p. 115). And in the comparisons of predictive values for binary diagnostic tests for paired designs, Leisenring et al. (2000) argued that although "the gains are small, the score statistic has consistently better size and power than a generalised Wald statistic" (p. 349).

To the best of our knowledge, the present paper is the first to propose the Wilson direct method and the Wilson indirect method to construct confidence intervals for the population $F_1$ score, to prove that the Wilson direct method does not suffer from both overshoot and degeneracy, and to compare the performance of both methods against the Clopper-Pearson method and the Wald method. Nonetheless, it is not without limitations. First, the present paper focused on constructing confidence intervals for a single population $F_1$ score but not for the difference between two or more population $F_1$ scores. Confidence intervals for the difference between two or more population $F_1$ scores should account for nonindependence between observations, especially if the confidence intervals are constructed using the same test set. Future research can investigate this, perhaps through the use of Generalised Estimating Equations (GEEs; e.g., Leisenring et al., 2000). Second, the present paper focused on analytical but not computational, and frequentist but not Bayesian, methods to construct either confidence or credible intervals for the population $F_1$ score. Future research can investigate this, perhaps building on current research focused on computational and/or Bayesian methods (Caelen, 2017; Flores et al., 2022; Goutte & Gaussier, 2005; Wang et al., 2015). Last, the present paper focused on the $F_1$ score but not the $F$-beta score. The $F$-beta score is the generalised form of the $F_1$ score in which the weights for both precision and recall are not constrained to be equal (Jurafsky & Martin, 2023; van Rijsbergen, 1975). Future research can investigate this, perhaps through the use of the multivariate Delta Method (Sen & Singer, 2018; Stone, 1995).

## 8    Conclusion

In conclusion, both the Wilson direct method and the Wilson indirect method are promising alternatives to both the Clopper-Pearson method and the Wald method to analytically constructing confidence intervals for the population $F_1$ score. Given the stochastic nature of evaluation in machine learning, it is recommended to construct and report confidence intervals when evaluating the performance of NLP algorithms.

**Table 1**

*Confusion Matrix for all Documents*

| Actual / Predicted | Positive | Negative |
|:---:|:---:|:---:|
| **Positive** | $p_{11}$ | $p_{10}$ |
| **Negative** | $p_{01}$ | $p_{00}$ |

**Table 2**

*Confusion Matrix for all Relevant Documents*

| Actual / Predicted | Positive | Negative |
|---|---|---|
| **Positive** | $\pi_{11}$ | $\pi_{10}$ |
| **Negative** | $\pi_{01}$ | — |

**Table 3**
*Coverage Probabilities*

| N | Scenario 1 | | | | Scenario 2 | | | | Scenario 3 | | | |
|---|---|---|---|---|---|---|---|---|---|---|---|---|
| | Clopper-Pearson | Wald | Wilson Direct | Wilson Indirect | Clopper-Pearson | Wald | Wilson Direct | Wilson Indirect | Clopper-Pearson | Wald | Wilson Direct | Wilson Indirect |
| 25 | 0.976 | 0.905 | 0.949 | 0.952 | 0.971 | 0.929 | 0.953 | 0.950 | 0.973 | 0.903 | 0.952 | 0.954 |
| 50 | 0.968 | 0.925 | 0.949 | 0.952 | 0.965 | 0.942 | 0.945 | 0.952 | 0.969 | 0.930 | 0.953 | 0.947 |
| 100 | 0.963 | 0.941 | 0.949 | 0.949 | 0.962 | 0.944 | 0.949 | 0.952 | 0.964 | 0.941 | 0.950 | 0.951 |
| 500 | 0.957 | 0.948 | 0.949 | 0.950 | 0.955 | 0.949 | 0.950 | 0.950 | 0.957 | 0.948 | 0.950 | 0.950 |
| 1000 | 0.955 | 0.949 | 0.950 | 0.950 | 0.954 | 0.949 | 0.950 | 0.950 | 0.955 | 0.949 | 0.950 | 0.950 |
| 5000 | 0.952 | 0.950 | 0.950 | 0.950 | 0.952 | 0.950 | 0.950 | 0.950 | 0.952 | 0.949 | 0.950 | 0.950 |

**Table 4**

*Expected Lengths*

| N | Scenario 1 | | | | Scenario 2 | | | | Scenario 3 | | | |
|---|---|---|---|---|---|---|---|---|---|---|---|---|
| | Clopper-Pearson | Wald | Wilson Direct | Wilson Indirect | Clopper-Pearson | Wald | Wilson Direct | Wilson Indirect | Clopper-Pearson | Wald | Wilson Direct | Wilson Indirect |
| 25 | 0.382 | 0.343 | 0.368 | 0.328 | 0.296 | 0.270 | 0.285 | 0.263 | 0.468 | 0.447 | 0.395 | 0.414 |
| 50 | 0.264 | 0.243 | 0.255 | 0.238 | 0.205 | 0.192 | 0.198 | 0.189 | 0.343 | 0.327 | 0.303 | 0.312 |
| 100 | 0.183 | 0.172 | 0.176 | 0.170 | 0.143 | 0.136 | 0.138 | 0.135 | 0.245 | 0.234 | 0.225 | 0.228 |
| 500 | 0.079 | 0.077 | 0.077 | 0.077 | 0.062 | 0.061 | 0.061 | 0.061 | 0.109 | 0.106 | 0.105 | 0.105 |
| 1000 | 0.055 | 0.054 | 0.054 | 0.054 | 0.044 | 0.043 | 0.043 | 0.043 | 0.076 | 0.075 | 0.075 | 0.075 |
| 5000 | 0.025 | 0.024 | 0.024 | 0.024 | 0.019 | 0.019 | 0.019 | 0.019 | 0.034 | 0.034 | 0.033 | 0.033 |

**Table 5**

*Overshoot Probabilities*

| N | Scenario 1 | | | | Scenario 2 | | | | Scenario 3 | | | |
|---|---|---|---|---|---|---|---|---|---|---|---|---|
| | Clopper-Pearson | Wald | Wilson Direct | Wilson Indirect | Clopper-Pearson | Wald | Wilson Direct | Wilson Indirect | Clopper-Pearson | Wald | Wilson Direct | Wilson Indirect |
| 25 | 0.000 | 0.231 | 0.000 | 0.000 | 0.000 | 0.020 | 0.000 | 0.000 | 0.000 | 0.200 | 0.000 | 0.000 |
| 50 | 0.000 | 0.006 | 0.000 | 0.000 | 0.000 | 0.000 | 0.000 | 0.000 | 0.000 | 0.031 | 0.000 | 0.000 |
| 100 | 0.000 | 0.000 | 0.000 | 0.000 | 0.000 | 0.000 | 0.000 | 0.000 | 0.000 | 0.000 | 0.000 | 0.000 |
| 500 | 0.000 | 0.000 | 0.000 | 0.000 | 0.000 | 0.000 | 0.000 | 0.000 | 0.000 | 0.000 | 0.000 | 0.000 |
| 1000 | 0.000 | 0.000 | 0.000 | 0.000 | 0.000 | 0.000 | 0.000 | 0.000 | 0.000 | 0.000 | 0.000 | 0.000 |
| 5000 | 0.000 | 0.000 | 0.000 | 0.000 | 0.000 | 0.000 | 0.000 | 0.000 | 0.000 | 0.000 | 0.000 | 0.000 |

**Table 6**

*Degeneracy Probabilities*

| N | Scenario 1 | | | | Scenario 2 | | | | Scenario 3 | | | |
|---|---|---|---|---|---|---|---|---|---|---|---|---|
| | Clopper-Pearson | Wald | Wilson Direct | Wilson Indirect | Clopper-Pearson | Wald | Wilson Direct | Wilson Indirect | Clopper-Pearson | Wald | Wilson Direct | Wilson Indirect |
| 25 | 0.000 | 0.004 | 0.000 | 0.000 | 0.000 | 0.000 | 0.000 | 0.000 | 0.000 | 0.013 | 0.000 | 0.000 |
| 50 | 0.000 | 0.000 | 0.000 | 0.000 | 0.000 | 0.000 | 0.000 | 0.000 | 0.000 | 0.000 | 0.000 | 0.000 |
| 100 | 0.000 | 0.000 | 0.000 | 0.000 | 0.000 | 0.000 | 0.000 | 0.000 | 0.000 | 0.000 | 0.000 | 0.000 |
| 500 | 0.000 | 0.000 | 0.000 | 0.000 | 0.000 | 0.000 | 0.000 | 0.000 | 0.000 | 0.000 | 0.000 | 0.000 |
| 1000 | 0.000 | 0.000 | 0.000 | 0.000 | 0.000 | 0.000 | 0.000 | 0.000 | 0.000 | 0.000 | 0.000 | 0.000 |
| 5000 | 0.000 | 0.000 | 0.000 | 0.000 | 0.000 | 0.000 | 0.000 | 0.000 | 0.000 | 0.000 | 0.000 | 0.000 |

**Figure 1**
*$F_1$ is a monotonic function of $F^*$ on the interval* [0,1]

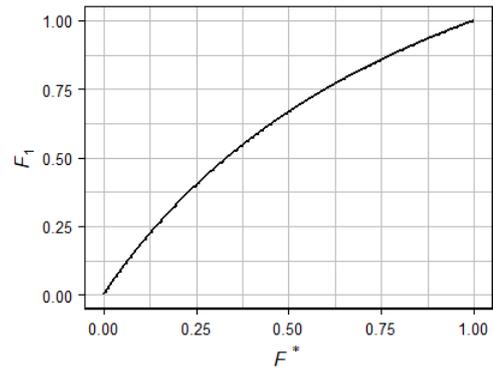

**Figure 2**

*Coverage Probabilities (Panel A) and Expected Lengths (Panel B) across all Simulation Conditions*

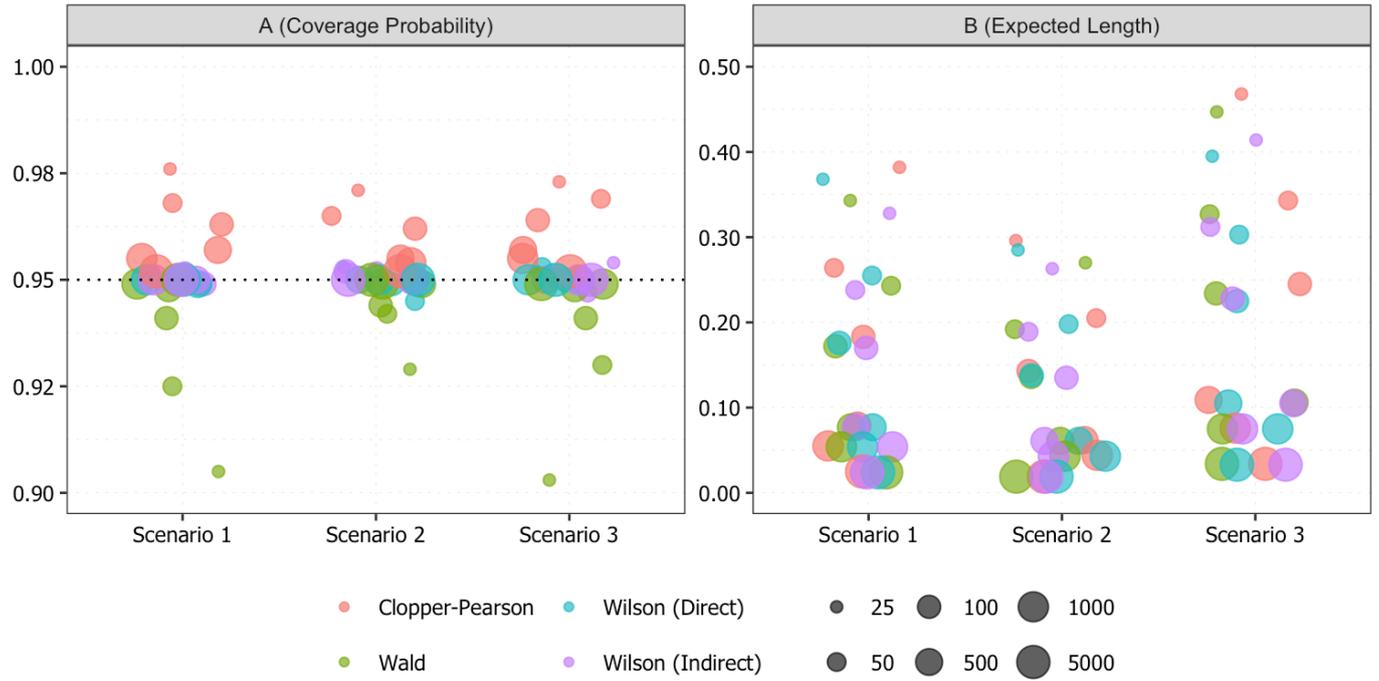

**Figure 3**

*A Comparison of Interval Lengths across Method (i.e., Wilson Direct vs. Wilson Indirect) and ν*

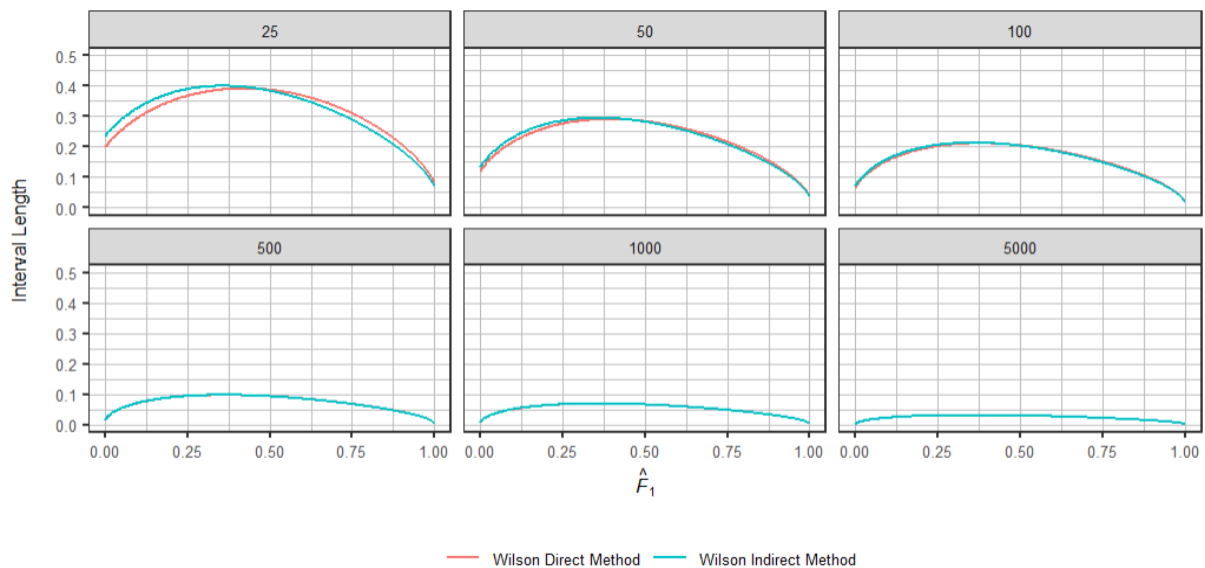